\documentstyle[preprint,aps]{revtex}
\newcommand{\beq}{\begin{equation}}
\newcommand{\eeq}{\end{equation}}
\newcommand{\beqa}{\begin{eqnarray}}
\newcommand{\eeqa}{\end{eqnarray}}
\newcommand{\ba}{\begin{array}}
\newcommand{\ea}{\end{array}}

\begin{document}
\draft

\widetext
\title {Bosons in a Toroidal Trap: Ground State and Vortices} 
\author {
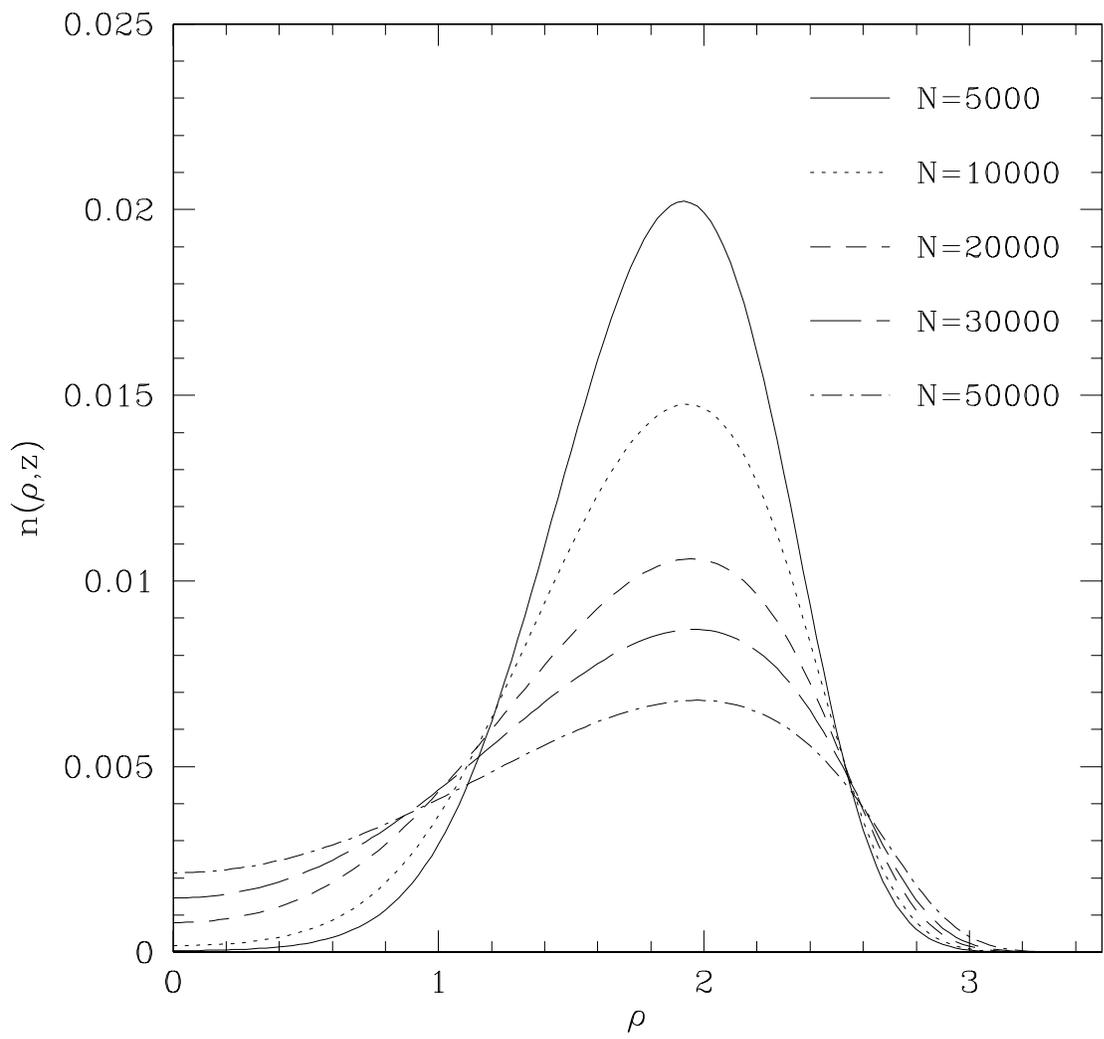
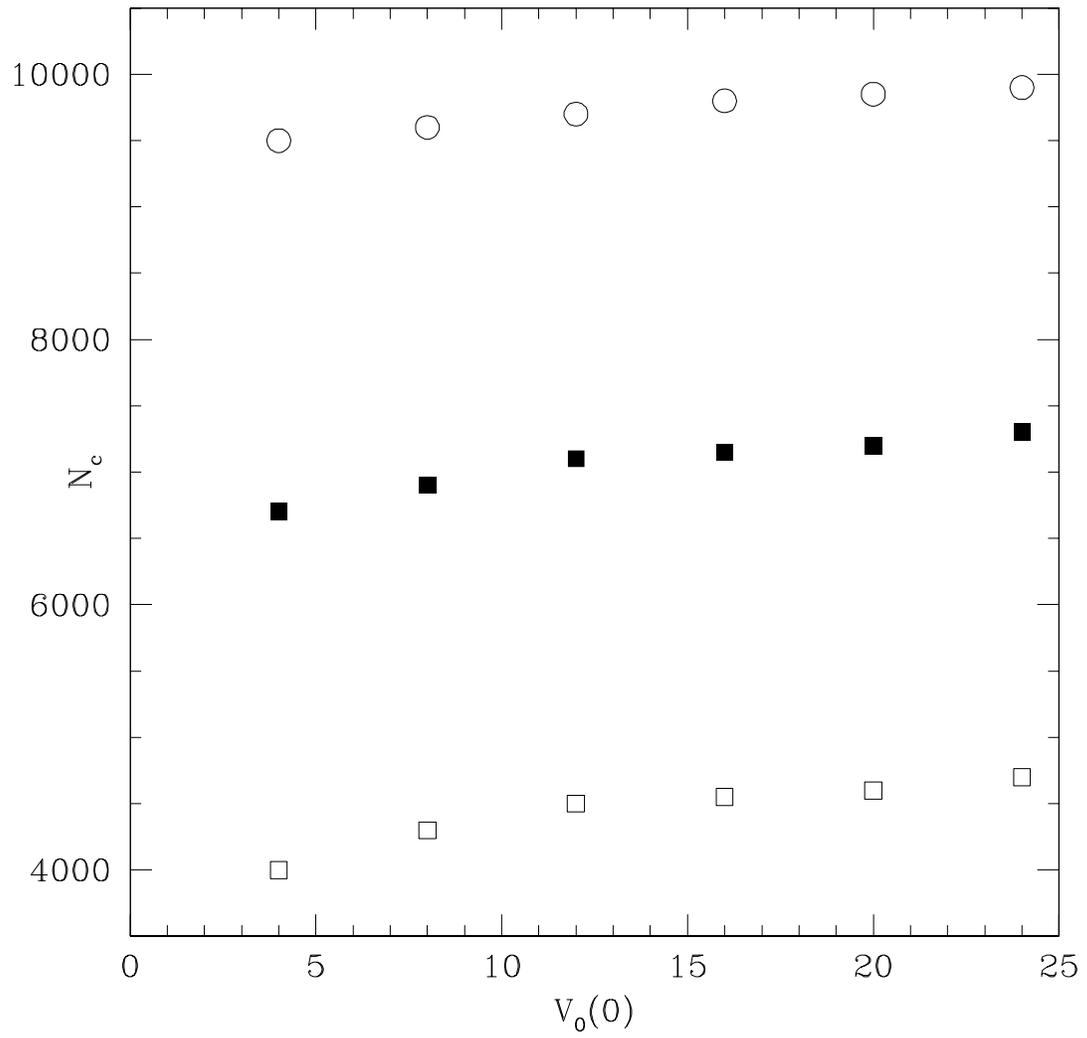
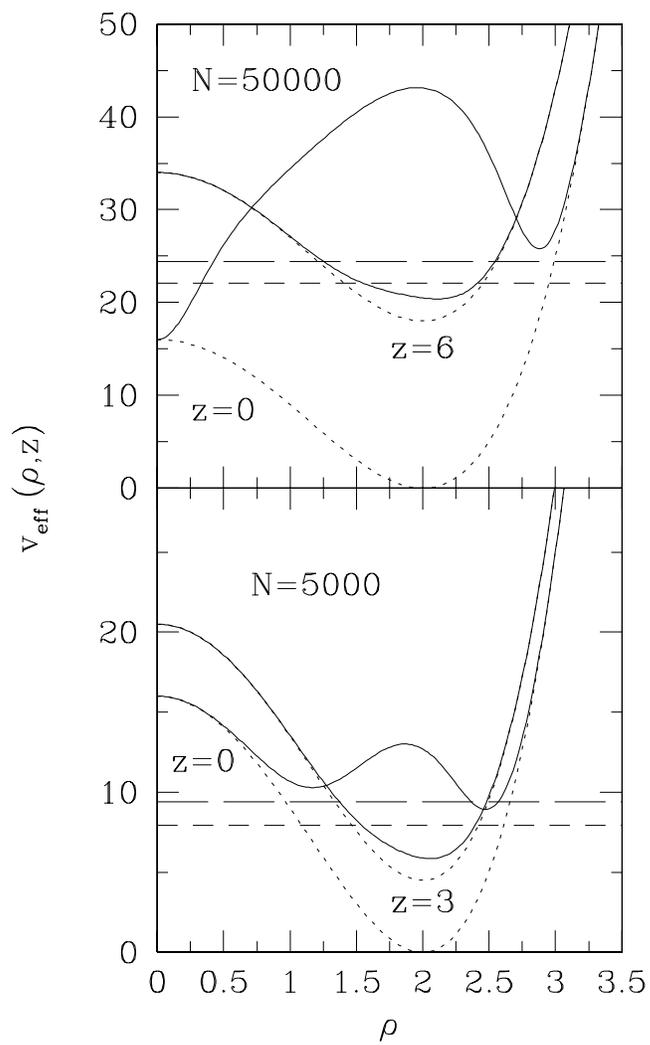
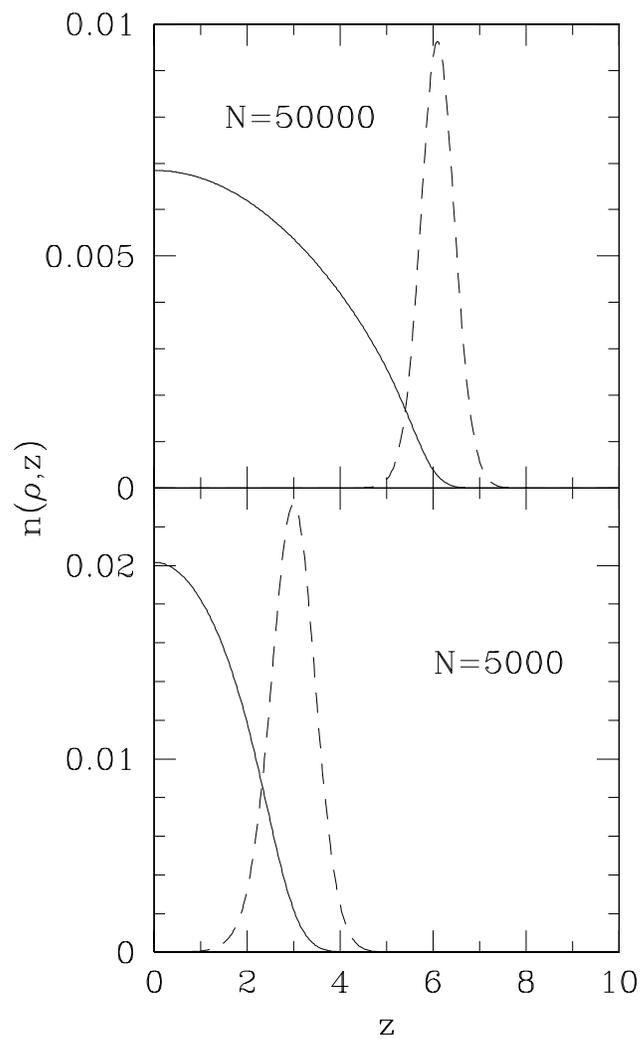
L. Salasnich$^{(*)(+)}$, A. Parola$^{(*)(++)}$ and L. Reatto$^{(*)(+)}$ }
\address{
$^{(*)}$ Istituto Nazionale per la Fisica della Materia, Unit\`a di Milano,\\
Via Celoria 16, 20133 Milano, Italy \\
$^{(+)}$ Dipartimento di Fisica, Universit\`a di Milano, \\
Via Celoria 16, 20133 Milano, Italy \\ 
$^{(++)}$ Dipartimento di Scienze Fisiche, Universit\`a dell'Insubria, \\ 
Via Lucini 3, 23100 Como, Italy }

\maketitle
\begin{abstract}
We study the Bose-Einstein condensate 
(BEC) in a 3-D toroidal Mexican hat trap. 
By changing the parameters of the potential, or the number of bosons, 
it is possible to modify strongly 
the density profile of the BEC. We consider the 
ground state properties for positive and negative scattering 
length and calculate the spectrum elementary excitations. 
We also discuss the macroscopic phase coherence 
and superfluidity of the BEC by analyzing 
vortex states and their stability. 
\end{abstract}
\narrowtext

\section{Introduction}
\par
Recent spectacular experiments 
with alkali vapors $^{87}Rb$, $^{23}Na$ and $^{7}Li$ 
confined in magnetic traps and cooled down to a temperature of the order of 
$100$ nK [1,2,3] have renewed the interest in the Bose-Einstein condensation. 
Theoretical studies of the Bose-Einstein condensate (BEC) in 
harmonic traps have been performed for the ground state [4,5,6,7], 
collective low-energy surface excitations [8,9] and vortex states 
[6]. The presence of vortex states is a signature of the 
macroscopic phase coherence of the system (the existence of a 
macroscopic quantum phase has been recently demonstrated [10]). Moreover, 
vortices are important to characterize the superfluid properties 
of Bose systems [11]. It has been found that the BEC in monotonically 
increasing potentials can not support stable vortices in the absence of 
an externally imposed rotation [12]. Instead, stable vortices 
of Bose condensates can be obtained in 1-D [13] and quasi-2D [14] 
toroidal traps: such Bose condensates are superfluid [11]. 
\par
In this paper we study a 3-D toroidal trap given by a quartic 
Mexican hat potential along the cylindrical radius and a harmonic potential 
along the $z$ axis. The resulting trapping 
potential is very flexible  and it is possible 
to modify considerably the density profile of the BEC 
by changing the parameters of the potential or the number of bosons. 
We analyze the ground state properties and the vortex stability 
of the condensate for both positive and negative scattering length and 
calculate also the spectrum of the Bogoliubov elementary excitations. 
In particular, we consider $^{87}Rb$ and $^{7}Li$ atoms. 
\par
The Gross-Pitaevskii 
energy functional [15] of the Bose-Einstein Condensate (BEC) reads: 
\beq
{E\over N} = \int d^3{\bf r} \;  
{\hbar^2\over 2m} |\nabla \Psi ({\bf r})|^2 
+ V_0({\bf r}) |\Psi ({\bf r})|^2 +{g N\over 2} |\Psi ({\bf r})|^4 \; ,
\eeq 
where $\Psi ({\bf r})$ is the wave function of the condensate
normalized to unity, $V_0({\bf r})$ 
is the external potential of the trap, and the interatomic potential 
is represented by a local pseudopotential so that $g={4\pi \hbar^2 a_s/m}$ 
is the scattering amplitude ($a_s$ is the s-wave scattering length). 
$N$ is the number of bosons of the condensate and $m$ is the atomic 
mass. The scattering length $a_s$ is supposed to be positive for 
$^{87}Rb$ and $^{23}Na$, but negative for $^{7}Li$. 
It means that for $^{87}Rb$ and $^{23}Na$ the interatomic 
interaction is repulsive while for $^{7}Li$ the atom-atom 
interaction is effectively attractive. 
The extremum condition for the energy functional 
gives the Gross-Pitaevskii (GP) equation [15]
\beq
\Big[ -{\hbar^2\over 2m} \nabla^2 
+ V_0({\bf r}) + g N |\Psi ({\bf r})|^2 \Big] \Psi ({\bf r}) 
= \mu \Psi ({\bf r}) \; ,
\eeq
where $\mu$ is the chemical potential. This equation has the form of 
a nonlinear stationary Schr\"odinger equation. 
\par
We study the BEC in an external Mexican hat potential 
with cylindrical symmetry, which is given by 
\beq
V_0({\bf r})={\lambda \over 4}
(\rho^2-\rho_0^2)^2+{m\omega_z^2 \over 2}z^2 \; , 
\eeq 
where $\rho=\sqrt{x^2+y^2}$ and $z$ 
are the cylindrical coordinates. 
This potential is harmonic along the $z$ axis and 
quartic along the cylindrical radius $\rho$. $V_0({\bf r})$ is minimum 
along the circle of radius $\rho = \rho_0$ at $z=0$ and $V_0({\bf r})$ 
has a local maximum at the origin in the $(x,y)$ plane. 
Small oscillations in the $(x,y)$ plane around $\rho_0$ have a 
frequency $\omega_{\perp}=\rho_0 (2\lambda /m)^{1/2}$.  
\par
First, let us consider the Thomas-Fermi (TF) approximation: 
i.e. neglect the kinetic energy. 
It is easy to show that the kinetic energy is negligible if 
$N>>(\hbar^2 / 2m)(\lambda \rho_0^2 + 
{m\omega_z^2/2})/\mu_0^2$, 
where $\mu_0=(2/ \pi^2 )(\lambda / 4)^{1/4}(m\omega_z^2 /2)^{1/4} 
g^{1/2}$ is the bare chemical potential. 
This condition is satisfied for $\lambda^2 \rho_0^8 >> 
16 \hbar^2(\lambda \rho_0^2 + m\omega_z^2/2)/(2 m)$. 
In the TF approximation we have 
\beq
\Psi ({\bf r}) = \Big[ {1\over g N}(\mu - V_0({\bf r})) \Big]^{1/2}
\Theta (\mu - V_0({\bf r}))  \; ,
\eeq
where $\Theta (x)$ is the step function. For our system 
we obtain that: 
a) the wave function has its maximum value at $\rho =\rho_0$ and $z=0$; 
b) for $\mu < \lambda \rho_0^4/4$ the wave function has a toroidal shape; 
c) for $\mu > \lambda \rho_0^4/4$ the wave function 
has a local minimum at $\rho=z=0$; 
d) the chemical potential scales as $\mu \sim \mu_0 N^{1/2}$.  
It is important to note that the TF approximation neglects 
tunneling effects: to include these processes, it is 
necessary to analyze the full GP problem. 

\section{Ground state properties and elementary excitations} 
\par
We perform the numerical minimization of the GP functional 
by using the steepest descent method [16]. 
It consists of projecting onto the minimum of the 
functional an initial trial state by propagating it in imaginary time. 
In practice one chooses a time step $\Delta \tau$ and iterates the equation
\beq
\Psi ({\bf r}, \tau +\Delta \tau) = \Psi ({\bf r}, \tau) 
- \Delta \tau {\hat H} \Psi ({\bf r}, \tau ) \; ,
\eeq
by normalizing $\Psi$ to $1$ at each iteration. 
\par
We discretize the space with a grid of points taking advantage of
the cylindrical symmetry of the problem.
At each time step the matrix elements entering 
the Hamiltonian are evaluated by means of finite--difference approximants. 
We use grids up to $200\times 200$ points verifying that the 
results do not depend on the discretization parameters.
The number of iterations in imaginary 
time depends on the degree of convergence required and the goodness 
of the initial trial wave function. We found that strict 
convergence criteria have to be required on the wave function 
in order to obtain accurate estimates of the wavefunction.
\par
In our calculations we use the $z$--harmonic oscillator units. 
We write $\rho_0$ in units $a_z=(\hbar / (m \omega_z))^{1/2} =1\;\mu$m, 
$\lambda$ in units $(\hbar \omega_z)a_z^{-4}= 0.477 \; (5.92)$ peV/$\mu$m$^4$ 
and the energy in units $\hbar \omega_z= 0.477 \; (5.92)$ peV 
for $^{87}Rb$ ($^{7}Li$). 
Moreover, we use the following values for the scattering length: 
$a_s=50 \;(-13)\; \AA$ for $^{87}Rb$ ($^{7}Li$) [1,3]. 
\par 
We have to distinguish two possibilities: positive or 
negative scattering length. In the case of positive scattering 
length we can control the density profile of the BEC by modifying 
the parameters of the potential and also the number of particles. 
In Figure 1 we show the ground state density profile of the $^{87}Rb$ 
condensate for several numbers of atoms. 
For small number of particles the condensate is essentially 
confined along the minimum of $V_0({\bf r})$, there is a very small 
probability of finding particles in the center of the trap so 
that the system is effectively multiply connected. As $N$ increases 
the center of the trap starts to fill up and the system becomes 
simply connected. The value of $N$ for which there is 
a crossover between the two regimes increases with 
the value of $\lambda$ and of $\rho_0$ and, within Thomas-Fermi
approximation,  scales like $\lambda^{3/2}\rho_0^8$. 
In Table 1 we show the energy per particle, the chemical potential 
and the average transverse and vertical size for the trapping potential 
characterized by the parameters $\rho_0=2$ and $\lambda=4$ in the 
$z$--harmonic oscillator units. 
As expected, the energy per particle and the chemical 
potential grow by increasing the number of particles 
but they do not scale as $N^{1/2}$ because, with this trapping potential, 
the TF approximation is valid for $N>>10^4$. It is instead 
interesting to observe that $\sqrt{<x^2>}=\sqrt{<y^2>}$ grows less than 
$\sqrt{<z^2>}$ due to the presence of a steep (quartic) potential 
along the transverse direction $\rho =\sqrt{x^2+y^2}$ 
and a softer (quadratic) barrier along the vertical direction $z$. 
\par 
In the case of negative scattering length, 
it is well known that for the BEC in harmonic potential 
there is a critical number of bosons $N_c$, 
beyond which there is the collapse of the wave function [7].  
We obtain the same qualitative behavior for the $^{7}Li$ condensate in our 
Mexican hat potential. However, in cylindrical symmetry, the collapse
occurs along the line which characterizes the minima of the external
potential, i.e. at $\rho=\rho_0$ and $z=0$. The numerical results
are shown in Figure 2. We notice that, 
for a fixed $\rho_0$, the critical number of bosons $N_c$ 
is only weakly dependent on the height of the barrier of the 
Mexican potential. These results suggest that we can not use 
toroidal traps to significantly enhance the 
metastability of the BEC with negative scattering length. 
\par 
To calculate the energy and wavefunction of the elementary excitations, 
one must solve the so--called Bogoliubov--de Gennes (BdG) equations [18,19]. 
The BdG equations can be obtained from the linearized 
time--dependent GP equation. Namely, one can look for zero angular momentum
solutions of the form 
\beq 
\Psi ({\bf r},t) = e^{-{i\over \hbar}\mu t} 
\Big[ \psi (\rho ,z) + u(\rho ,z) e^{-i \omega t} 
+v^*(\rho ,z) e^{i \omega t} \Big] \; ,
\eeq
corresponding to small oscillations of the wavefunction 
around the ground state solution $\psi$.
By keeping terms linear in the complex functions $u$ and $v$, one finds 
the following BdG equations 
$$
\Big[ -{\hbar^2\over 2m} 
\Big({\partial^2 \over \partial \rho^2} + 
{1\over \rho} {\partial \over \partial \rho} + 
{\partial^2 \over \partial z^2} \Big) 
+ V_0(\rho,z) - \mu + 2g N |\psi (\rho,z)|^2 \Big] u(\rho,z) + 
$$
\beq
+ g N |\psi (\rho,z)|^2 v(\rho,z) = \hbar \omega \; u(\rho,z) \; , 
\eeq
$$
\Big[ -{\hbar^2\over 2m} 
\Big({\partial^2 \over \partial \rho^2} + 
{1\over \rho} {\partial \over \partial \rho} + 
{\partial^2 \over \partial z^2} \Big) 
+ V_0(\rho,z) - \mu + 2g N |\psi (\rho,z)|^2 \Big] v(\rho,z) + 
$$
\beq 
+ g N |\psi (\rho,z)|^2 u(\rho,z)= - \hbar \omega \; v(\rho,z) \; . 
\eeq
The BdG equations allow one to calculate the eigenfrequencies $\omega$ 
and hence the energies $\hbar \omega$ of the elementary 
excitations. This procedure is equivalent to the 
diagonalization of the N--body Hamiltonian of the system in the Bogoliubov 
approximation [17]. The excitations can be classified according to 
their parity with respect to the symmetry $z\to -z$.
\par 
We have solved the two BdG eigenvalue equations by finite--difference 
discretization with a lattice of $40\times 40$ points 
in the $(\rho,z)$ plane. In this way, the eigenvalue problem reduces to the 
diagonalization of a $3200\times 3200$ real matrix. 
We have tested our program in simple models by comparing numerical
results with the analytical solution and 
verified that a $40 \times 40$ mesh already gives reliable
results for the lowest part of the spectrum. In Table 3 we show the 
lowest elementary excitations of the Bogoliubov spectrum for 
the ground state of the system. One observes the presence of an odd
collective excitation at energy quite close to $\hbar \omega =1$ (in units 
$\hbar \omega_z$). This mode is related to the oscillations of the 
center of mass of the condensate which, due to the harmonic 
confinement along the $z$--axis, is an exact eigenmode of the problem 
characterized by the 
frequency $\omega_z$, independently of the strength of the interaction.
For large $N$ the lowest elementary excitations saturate, 
suggesting that the Thomas--Fermi asymptotic limit is reached. 

In the case of negative scattering length we verified that, quite close
to the critical number of bosons $N_c$, an even mode 
softens driving the transition towards a collapsed state.

\section{Vortices and their metastability}
\par
Let us consider states having a vortex line 
along the $z$ axis and all bosons flowing around it with 
quantized circulation. The observation of these states would be 
a signature of macroscopic phase coherence of trapped BEC. 
The axially symmetric condensate wave function can be written as 
\beq
\Psi_k ({\bf r}) = \psi_k (\rho , z) \; e^{i k \theta} \; ,
\eeq
where $\theta$ is the angle around the $z$ axis and $k$ is the integer 
quantum number of circulation. The resulting GP functional (1),
representing the energy per particle,
can be written in terms of $\psi_k({\bf r})$ by taking advantage of the
cylindrical symmetry of the problem:
$$ 
{E\over N} = 
\int \rho \; d\rho \; dz \; d\theta \; {\hbar^2\over 2m} 
\Big[ 
|{\partial \psi_k (\rho,z)\over \partial\rho}|^2 + 
|{\partial \psi_k (\rho,z)\over \partial z}|^2 \Big]+ 
$$
\beq  
+\Big[{\hbar^2 k^2 \over 2m\rho^2} 
+ V_0(\rho,z) \Big] \left\vert\psi_k (\rho,z)\right\vert^2 
+{gN\over 2} |\psi_k (\rho,z)|^4  \; .
\eeq
Due to the presence of the centrifugal term, the solution 
of this equation for $k\neq 0$ has to vanish on the $z$ axis 
providing a signature of the vortex state.
\par
Vortex states are important to characterize the macroscopic 
quantum phase coherence and also superfluid properties 
of Bose systems [11]. It is easy to calculate the critical frequency 
$\Omega_c$ at which a vortex can be produced. 
One has to compare the energy of a vortex state in a frame rotating 
with angular frequency $\Omega$, that is $E-\Omega L_z$, with 
the energy of the ground state with no vortices. Since the 
angular momentum per particle is $\hbar k$, the critical 
frequency is given by $\hbar \Omega_c =(E_k/N - E_0/N)/k$, 
where $E_k/N$ is the energy per 
particle of the vortex with quantum number $k$. In 
Table 2 we show some results for vortices of $^{87}Rb$. 
The critical frequency turns out to increase slightly
with the number of atoms. This corresponds to a moderate
lowering of the momentum of inertia per unit mass of the condensate
when $N$ grows.
\par
For $^{7}Li$ we calculate the critical number $N_c$ of bosons for 
which there is the collapse of the vortex wave function. We find that $N_c$ 
has a rather weak dependence on the quantum number of circulation $k$. 
Note that, in the case of a harmonic external potential, there is 
an enhancement of $N_c$ by increasing $k$ because 
in that case rotation strongly reduces the density in the 
neighborhood of the origin, where the external potential 
has its minimum [6]. 
\par 
Once a vortex has been produced, the BEC is superfluid if 
the circulating flow persists, in a metastable state, 
in the absence of an externally imposed rotation [11]. 
As discussed previously, vortex solutions centered in 
harmonic traps have been found [6], but such states turn out 
to be unstable to single particle excitations out of the condensate.
To study the metastability of the vortex we first analyze the 
following Hartree-Fock equation [12] 
\beq
\Big[ -{\hbar^2\over 2m} \nabla^2 
+ V_0({\bf r}) + 2g N |\psi_k ({\bf r})|^2 \Big] \phi ({\bf r}) 
= \epsilon \phi ({\bf r}) \; , 
\eeq
which describes, in the weak coupling limit, 
one particle transferred from the vortex state 
$\Psi_k({\bf r})$ to an orthogonal single-particle state $\phi ({\bf r})$. 
Quasiparticle motion is governed by an effective Hartree potential 
$v_{eff}({\bf r})=V_0({\bf r}) + 2g |\psi_k ({\bf r})|^2$, 
which combines the effects 
of the trap with a mean repulsion by the condensate. 
Figure 3 shows $v_{eff}(\rho,z)$ for $N=5000$ and $50000$.
The repulsion induced by the underlying condensate is quite
evident near $\rho=\rho_0$.
Let $\mu_k$ be the chemical potential of the vortex state characterized
by a circulation quantum number $k$, 
then the vortex is metastable if $\epsilon > \mu_k$ and unstable 
if $\epsilon < \mu_k$ [12]. As shown in Table 2, 
for our 3-D system all the studied vortices are metastable 
and so the BEC can support persistent currents, thus it is superfluid. 
Contrary to what may be inferred by means of semiclassical arguments [12]
the wave function describing the excitation $\phi({\bf r})$ is not 
localized near the symmetry axis even for rather large numbers of atoms. 
A bound state at $\rho=z=0$ should
pay a large kinetic energy cost due to the strong localization
of the particle induced by the effective potential. Instead, it is more
convenient to place the excited particle on top of the Bose condensate
i.e. at $\rho=\rho_0$ and $z\ne 0$ as shown in Figure 4. 
\par
It is well known that the Hartre--Fock approximation describes only 
single particle excitations [17]. To have the complete spectrum, 
including collective excitations, one must solve the BdG equations [18,19]. 
One must look for solutions of the form 
\beq 
\Psi_k ({\bf r},t) = e^{-{i\over \hbar}\mu_k t} 
\Big[ \psi_k (\rho ,z) e^{i k \theta} + u(\rho ,z) 
e^{i (k+q) \theta} e^{-i \omega t} 
+v^*(\rho ,z) e^{i (k-q) \theta} e^{i \omega t} \Big] \; .
\eeq
Here $q$ represents the quantum number of circulation of the elementary
excitation. We have solved the two BdG eigenvalue equations 
by finite--difference discretization using the same method described
in Section II. We have checked that a $40 \times 40$ mesh gives 
the correct excitation energies within Hartree-Fock approximation.
Therefore, for the purpose of determining the stability of the
vortex state, this rather coarse mesh is sufficiently accurate.
The results are shown in Table 4: 
The lowest Bogoliubov excitation is positive and always lower than 
the lowest Hartree--Fock one. 
Moreover by increasing the number of particles their difference 
increases as expected for collective excitations. 
We have also verified that vortex states become unstable by strongly reducing 
either density (down to about one hundred bosons in our model trap) 
or scattering length. 
\par 
Therefore, the behavior of the 3-D trap we have analyzed closely resembles 
the simplified 1-D model studied in Ref. [13] which represents the limit of 
deep trapping potential. Also in that case Bogoliubov approximation has 
been used to evaluate the spectrum of elementary excitations showing 
that vortices are stabilized by strong repulsive interparticle 
interactions (or equivalently by high density). The 1-D model, however, 
should be taken with caution because other branches of low energy collective
excitations are present in such low-dimensional systems [20]. 

\section{Conclusions}
\par
We have studied the Bose-Einstein condensate in a 
3-D toroidal trap given by a quartic 
Mexican hat potential along the cylindrical radius and a harmonic potential 
along the $z$ axis. We have shown that 
it is possible to modify strongly the density profile of the condensate 
by changing the parameters of the potential or the number of bosons. 
The properties of the condensate and its elementary excitations 
have been analyzed for both 
positive and negative scattering length by considering 
$^{87}Rb$ and $^{7}Li$ atoms. 
For $^{7}Li$, which has negative scattering length, we 
have calculated the critical number of atoms for which there is 
the collapse of the wave function. The results have shown that a 
toroidal trap does not enhance the metastability of the 
ground state of the condensate. On the other hand, 
in the case of a harmonic external potential, 
we have recently shown [21,22] that, when a realistic non local (finite range) 
effective interaction is taken into account, a new stable 
branch of Bose condensate appears for $^7Li$ at higher density. 
Presumably a similar state can be found also in presence 
of a toroidal external trap for a sufficiently large number of 
particles when non locality effects are included. 
\par
A superfluid is characterized by the presence of persistent currents
in the absence of an externally imposed rotation. In order to
investigate this peculiar sign of the macroscopic phase 
coherence of the condensate, we have also studied vortex states. 
Our results suggest that vortices can support persistent 
currents in 3-D toroidal traps with fairly large numbers of atoms. 
This feature essentially depends on the toroidal geometry 
of the trap and should be independent on other details of the 
confining potential. 

\section*{Acknowledgements}
\par
This work has been supported by INFM under the Research Advanced Project (PRA) 
on "Bose-Einstein Condensation". 

\newpage

\section*{References}

\begin{description}

\item{\ [1]} M.H. Anderson, J.R. Ensher, M.R. Matthews, C.E. Wieman, 
and E.A. Cornell, Science {\bf 269}, 189 (1995). 

\item{\ [2]} K.B. Davis, M.O. Mewes, M.R. Andrews, N.J. van Druten, 
D.S. Drufee, D.M. Kurn, and W. Ketterle, Phys. Rev. Lett. {\bf 75}, 
3969 (1995).

\item{\ [3]} C.C. Bradley, C.A. Sackett, J.J. Tollet, and R.G. Hulet, 
Phys. Rev. Lett. {\bf 75}, 1687 (1995). 

\item{\ [4]} M. Edwards and K. Burnett, Phys. Rev. A {\bf 51}, 1382 (1995).

\item{\ [5]} M. Lewenstein and L. You, Phys. Rev. A {\bf 53}, 909 (1996).

\item{\ [6]} F. Dalfovo and S. Stringari, Phys. Rev. A {\bf 53}, 2477 (1996).

\item{\ [7]} R.J. Dodd {\it et al}, Phys. Rev. A  {\bf 54}, 661 (1996); 
G. Baym and C.J. Pethick, Phys. Rev. Lett. {\bf 76}, 6 (1996).

\item{\ [8]} S. Stringari, Phys. Rev. Lett. {\bf 77}, 2360 (1996).

\item{\ [9]} A. Smerzi and S. Fantoni, Phys. Rev. Lett. {\bf 78}, 3589 
(1997). 

\item{\ [10]} M.R. Andrews, C.G. Townsend, H.J. Miesner, 
D.S. Drufee, D.M. Kurn, and W. Ketterle, Science {\bf 275}, 
637 (1997).

\item{\ [11]} A.J. Leggett, in {\it Low Temperature Physics}, 
Lecture Notes in Physics, vol. 394, pp. 1-91, Ed. M.J.R. Hoch and 
R.H. Lemmer (Springer, Berlin, 1991). 

\item{\ [12]} D.S. Rokhsar, Phys. Rev. Lett. {\bf 79}, 2164 (1997).

\item{\ [13]} D.S. Rokhsar, ``Dilute Bose gas in a torus: 
vortices and persistent currents'', cond-mat/9709212. 

\item{\ [14]} M. Benakli, S. Raghavan, A. Smerzi, S. Fantoni and R. 
Shenoy, ``Macroscopic Angular Momentum States of Bose-Einstein 
Condensates in Toroidal Traps", cond-mat/9711295. 

\item{\ [15]} E.P. Gross, Nuovo Cimento {\bf 20}, 454 (1961); 
L.P. Pitaevskii, Sov. Phys. JETP {\bf 13}, 451 (1961).

\item{\ [16]} S. Koonin and C.D. Meredith, {\it Computational Physics} 
(New York, 1990).

\item{\ [17]} F. Dalfovo, S. Giorgini, M. Guilleumas, L.P. Pitaevskii  
and S. Stringari, Phys. Rev. A {\bf 56}, 3804 (1997). 

\item{\ [18]} A.L. Fetter, Phys. Rev. A {\bf 53}, 4245 (1996). 

\item{\ [19]} D.A.W. Hutchinson, E. Zeremba, A. Griffin, 
Phys. Rev. Lett. {\bf 78}, 1842 (1997).  

\item{\ [20]} E.H. Lieb and W. Lininger, Phys. Rev. {\bf 130}, 
1616(1963).

\item{\ [21]} A. Parola, L. Salasnich and L. Reatto, Phys. Rev. A 
{\bf 57}, 3180 (1998).

\item{\ [22]} L. Reatto, A. Parola and L. Salasnich, J. Low Temp. Phys. 
{\bf 113}, N. 3 (1998). 

\end{description}

\newpage

\section*{Figure Captions} 

\vskip 1. truecm

{\bf Figure 1}. 
Particle probability density in the ground state of $^{87}Rb$ atoms 
as a function of the cylindrical radius at $z=0$ (symmetry plane).
The curves correspond to different numbers of atoms: from $5000$ to 
$50000$. Parameters of the external potential: $\rho_0=2$ and $\lambda =4$. 
Lengths are in units of $a_z =1\; \mu$m and $\lambda$ in units of 
$(\hbar \omega_z)a_z^{-4}=0.477$ peV/$\mu$m$^4$. 

\vskip 1. truecm

{\bf Figure 2}. Critical number $N_c$ of $^{7}Li$ atoms {\it versus} 
the potential barrier at the origin: $V_0(0)=\lambda \rho_0^4/4$. 
Open squares: $\rho_0=2$; full squares: $\rho_0=3$; 
open circles: $\rho_0=4$. Energy is in units of 
$\hbar \omega_z=5.92$ peV ($\omega_z=9.03$ kHz) and length 
in units of $a_z=1\; \mu$m. 

\vskip 1. truecm

{\bf Figure 3}. Effective potential $v_{eff}(\rho,z)$ appearing in 
the eigenvalue equation for the single particle excitation Eq. (8).
Two sections at $z=0$ and $z=3$ ($z=6$) are shown for 
$N=5000$ ($N=50000$) atoms in panel a (b). $z=0$ corresponds to the
symmetry plane. The dotted line represents the external potential.
The chemical potential of the vortex state is
marked by a short dashed line, the excitation energy by a long dashed
line. Parameters of the external potential: $\rho_0=2$ and $\lambda =4$. 
Units as in Fig. 1 with $\hbar \omega_z=0.477$ peV 
($\omega_z = 0.729$ kHz). 

\vskip 1. truecm

{\bf Figure 4}. 
Particle probability density of the $k=1$ vortex state
(solid line) and square of the excitation wave function (dashed line)
at the radial distance $\rho=\rho_0=2$, i.e. where both the 
wave functions peak. Curves are for 
$N=5000$ ($N=50000$) atoms in panel a (b). Units and parameters 
as in Fig. 3. 

\newpage

%\begin{table}
\begin{center}
\begin{tabular}{|ccccc|} \hline
$N$ & $E/N$ & $\mu$ & $\sqrt{<\rho^2>}$ & $\sqrt{<z^2>}$ \\ 
\hline
$5000$  & $5.85$  & $7.71$  & $1.96$ & $1.41$  \\
$10000$ & $7.45$  & $10.26$ & $1.97$ & $1.70$  \\ 
$20000$ & $9.84$  & $14.00$ & $1.97$ & $2.05$  \\
$30000$ & $11.73$ & $16.94$ & $1.97$ & $2.29$  \\
$40000$ & $13.36$ & $19.47$ & $1.98$ & $2.48$  \\
$50000$ & $14.81$ & $21.74$ & $1.99$ & $2.63$  \\
\hline 
\end{tabular} 
\end{center} 
%\end{table} 

\vskip 0.3 truecm
{\bf Table 1}. Ground state of $^{87}Rb$ atoms in the toroidal 
trap with $\rho_0 =2$ and $\lambda=4$. 
Chemical potential and energy are in units of $\hbar \omega_z=
0.477$ peV ($\omega_z=0.729$ kHz). Lengths are in units of $a_z=1\;\mu$m. 

\vskip 1. truecm

%\begin{table}
\begin{center}
\begin{tabular}{|ccccc|} \hline
$N$ & $E_1/N$ & $\mu_1$ & $\epsilon$ & $\hbar \Omega_c$ \\ 
\hline
$5000$  & $6.00$  & $7.87$  & $9.56$   & $0.15$  \\ 
$10000$ & $7.61$  & $10.44$ & $12.46$  & $0.16$  \\
$20000$ & $10.02$ & $14.22$ & $16.60$  & $0.18$  \\
$30000$ & $11.93$ & $17.20$ & $19.80$  & $0.20$  \\
$40000$ & $13.57$ & $19.75$ & $22.54$  & $0.21$  \\ 
$50000$ & $15.04$ & $22.04$ & $25.09$  & $0.23$  \\ 
\hline 
\end{tabular} 
\end{center} 
%\end{table} 

\vskip 0.3 truecm 
{\bf Table 2}. Vortex states and excitation energies of $^{87}Rb$ atoms 
with $k=1$ in the toroidal trap with $\rho_0 =2$ and $\lambda=4$ 
within Hartree-Fock approximation. Units as in Tab. 1. 

\vskip 1. truecm

%\begin{table} 
\begin{center}
\begin{tabular}{|ccccc|} \hline
$N$ & $\hbar\omega_1$ & $\hbar\omega_2$ & $\hbar\omega_3$ & $\hbar\omega_4$ \\ 
\hline 
$1$     & $1.00$ & $1.98$ & $2.97$ & $3.96$ \\
$5000$  & $1.00$ & $1.70$ & $2.43$ & $3.19$ \\ 
$10000$ & $1.00$ & $1.68$ & $2.37$ & $3.08$ \\
$20000$ & $1.00$ & $1.66$ & $2.32$ & $3.00$ \\
$30000$ & $1.00$ & $1.66$ & $2.30$ & $2.96$ \\ 
$40000$ & $1.00$ & $1.66$ & $2.30$ & $2.95$ \\
$50000$ & $1.00$ & $1.66$ & $2.30$ & $2.95$ \\ 
\hline 
\end{tabular}
\end{center} 
%\end{table}

\vskip 0.3 truecm 
{\bf Table 3}. Lowest elementary excitations of the 
Bogoliubov spectrum for the ground state of $^{87}Rb$ atoms in the toroidal 
trap with $\rho_0 =2$ and $\lambda=4$. 
Units as in Tab. 1. 

\vskip 1. truecm

%\begin{table} 
\begin{center}
\begin{tabular}{|ccc|} \hline
$N$ & $\hbar \omega$ & $\epsilon - \mu_1$ \\ 
\hline 
$5000$  & $1.22$ & $1.69$ \\ 
$10000$ & $1.48$ & $2.02$ \\
$20000$ & $1.73$ & $2.38$ \\
$30000$ & $1.88$ & $2.60$ \\ 
$40000$ & $1.99$ & $2.79$ \\
$50000$ & $2.08$ & $3.05$ \\ 
\hline 
\end{tabular}
\end{center} 
%\end{table}

\vskip 0.3 truecm 
{\bf Table 4}. Bogoliubov {\it vs} Hartree--Fock lowest 
elementary excitation for a vortex state of $^{87}Rb$ atoms 
with $k=1$ in the toroidal trap with $\rho_0 =2$ and $\lambda=4$. 
Units as in Tab. 1. 

\end{document}